\documentclass[runningheads]{llncs}
\usepackage[T1]{fontenc}

\usepackage{graphicx}
\usepackage{url}
\usepackage{verbatim}
\usepackage{hyperref}
\usepackage{enumitem}
\usepackage{float}
\usepackage{tikz}
\usetikzlibrary{positioning, shapes.geometric, arrows}
\usepackage{microtype}

\begin{document}
\title{Proposal of an AI-Based Support Assistant \\ for the ALICE-FIT Detector Setup at CERN}
\titlerunning{AI-Based Support Assistant for the ALICE-FIT Detector Setup at CERN}
%

\author{
Ignacy Mermer\inst{1}\orcidID{0009-0009-7777-4008}, Jakub Muszyński\inst{1}\orcidID{0009-0000-2797-6044}, Jakub Możaryn\inst{1}\orcidID{0000-0002-2677-0113}, Krystian Rosłon\inst{1}\orcidID{0000-0002-6732-2915}, } 
\authorrunning{I.Mermer, et al.}
%
\institute{
Warsaw University of Technology, Warsaw, Poland\\
\email{krystian.roslon@pw.edu.pl}}
\maketitle              
\begin{abstract}

We propose an AI-based assistant designed to support the ALICE Fast Interaction Trigger (FIT) detector operators at CERN. The assistant helps diagnose and resolve operational issues in the Detector Control System (DCS), where decisions must often be made quickly and with incomplete information. By combining Large Language Models (LLMs) with a controlled Retrieval-Augmented Generation (RAG) pipeline, the system can generate context-aware suggestions based on verified ALICE-FIT documentation and problems that have appeared in the past.

\end{abstract}
\section{Introduction and Objectives}
The ALICE (A Large Ion Collider Experiment) at CERN explores the properties of strongly interacting matter in heavy-ion collisions, providing fundamental insights into the quark–gluon plasma \cite{Acharya2024}\cite{Slupecki:2022fch}. Among its subsystems, the Fast Interaction Trigger plays a critical and enabling role: it provides collision timing, luminosity monitoring, and the primary interaction trigger for the entire experiment \cite{Slupecki:2022fch}\cite{Roslon:2025lxb}. Without a fully operational FIT, in particular FIT-FT0, ALICE cannot acquire physics data. Ensuring FIT’s continuous, error-free performance is therefore essential for every stage of data taking at the Large Hadron Collider (LHC).

In routine operation, the ALICE experiment follows a well-defined workflow. When the LHC delivers stable beams, the central Run Control configures all detector subsystems into a coherent global state and supervises data taking. Each detector (including FIT) is managed by its own DCS, which exposes a high-level Finite State Machine (FSM) to the experiment and encapsulates detailed hardware procedures. Shifters in the control room interact, 24 hours a day, primarily with these FSM states and summary views; expert teams define and maintain the underlying configuration, calibration, and safety logic. The overall philosophy is to automate as much as is safely possible, while keeping humans in the loop for non-standard situations and high-impact decisions.

From the point of view of a FIT operator, this results in a hierarchical chain of responsibilities. At the lowest level, front-end boards, High-Voltage channels, and crates expose raw status and configuration registers. Above them, FRED (Front-End Device)~\cite{Roslon:2025zee} and ALF/IPbus -- ALF~(Alice Low-level Frontend))~\cite{Roslon:2025zee} aggregate this information and provide services to the Supervisory Control And Data Acquisition (SCADA) system. WinCC~OA presents alarm panels, trend plots, and FSM panels, which the shifter monitors while following written procedures. As a concrete example, consider a situation in which a subset of FIT channels reports intermittent data-quality degradation: the shifter observes an alarm in WinCC~OA, checks the FSM state of FIT, resolves typical problems on the side of Front-End Electronics (FEE) or High Voltage (HV) based on prepared instructions. Diagnosing whether this is caused by a transient communication glitch, a misconfigured FEE parameters, or a~genuine hardware problem requires correlating information from several tools under time pressure.

Operating large detector systems such as FIT is never free from occasional faults or inconsistencies. These range from low-level communication errors to mismatched configuration states that may interrupt data taking. While most are automatically mitigated by interlocks, many still require manual diagnosis within the Detector Control System, which serves as the central operational layer linking hardware states, monitoring feedback, and human decision-making \cite{Roslon:2025lxb}\cite{Roslon:2025zee}\cite{Roslon2025AUW}\cite{Roslon:2025tym}\cite{Roslon:2025buc}.

However, as the DCS evolves and the amount of telemetry, procedural knowledge, and operational interdependencies grows, manual problem-solving becomes increasingly demanding. Operators must integrate information from diverse sources --- such as eLogs, Bookkeeping entries, and documentation --- under time pressure and safety constraints \cite{Roslon2025AUW}\cite{Roslon:2025tym}. To support this complex decision-making process, the project introduces an AI-based Support Assistant for the FIT DCS. The assistant is designed to provide context-aware, knowledge-grounded, and safe guidance, combining the reasoning capabilities of Large Language Models with structured retrieval from verified ALICE documentation. Ultimately, this approach aims to improve operational efficiency and reliability in FIT and to serve as a scalable model for AI-assisted control in high-energy physics experiments.

The concrete objectives of the proposed assistant are:
\begin{itemize}[leftmargin=*,nosep]
    \item to consolidate disparate operational knowledge (procedures, incident reports, configuration notes) into a curated, queryable knowledge base;
    \item to provide conversational, step-by-step diagnostic support that mirrors expert reasoning while remaining traceable to underlying documentation;
    \item to interface with the FIT DCS stack in a safe, auditable way, initially in read-only mode and later with carefully gated \emph{Quick Commands};
    \item to evaluate the assistant not only in terms of technical KPIs (accuracy, latency, workload reduction) but also in terms of operator experience, perceived workload, and trust in the system.
\end{itemize}

%
\section{Literature Review and Methods}

Operating a large High Energy Physics (HEP) detector entails time-critical decisions under codified procedures. In ALICE, detector control relies on \emph{WinCC~OA} with the CERN \emph{JCOP framework}; operational knowledge accumulates in the eLog and the new \emph{Bookkeeping} platform; and the Run~3 \emph{O$^2$} stack defines the online/offline services an assistant must respect when consuming status and metadata \cite{CERN_JCOP_Service,ALICE_Bookkeeping_2024,EulisseRohr_O2_2024}. While AI/ML has delivered substantial impact across LHC workflows \cite{Guest:2018yhq,Radovic:2018dip,Cerri:2019qrb} and recent position papers outline the promise of LLMs in HEP \cite{Kasieczka:2023ell}, bringing these gains to \emph{control-room support} requires methods that are knowledge-grounded and safety- and authorization-aware. We therefore adopt Retrieval-Augmented Generation to reduce hallucinations by conditioning outputs on curated corpora (procedures, JCOP/DCS documentation, incident reports, selected O$^2$ docs) \cite{Lewis:2020gnp}; beyond baseline RAG, \emph{Self-RAG} learns \emph{when} to retrieve and \emph{what} to cite, improving factuality and long-form citation quality in technical domains \cite{Asai_SelfRAG_2023}. In our deployment, the index prioritizes ALICE manuals, Bookkeeping entries and vetted eLog excerpts, and the assistant attaches retrieved snippets and provenance to answers for auditability. Because the assistant must \emph{act} as well as answer, we employ tool-using agent designs in which models decide when to call external tools/APIs and how to incorporate results \cite{Schick:2023toolformer}; specifically, we instantiate \emph{ReAct}, interleaving reasoning, action and observation, for transparent, stepwise diagnostics that mirror operator workflows \cite{Yao:2023react}. Tool access is read-only by default, with controlled escalation (write actions) gated by operator confirmation and role checks. Safety is first-class: we encode behavioral constraints using principles from \emph{Constitutional AI} and complement them with programmable runtime guardrails \cite{Bai_ConstitutionalAI_2022,Rebedea_NeMoGuardrails_2023}; because retrieval widens the attack surface, we follow OWASP GenAI guidance to mitigate prompt-injection and indirect manipulation via input/output filtering and retrieval-layer sanitization \cite{OWASP_LLM_Top10_2025}. All privileged operations traverse a service layer enforcing \emph{role-based access control} (RBAC) with full audit logging, anchored in the NIST/ANSI model \cite{Sandhu_RBAC_1996}.

Thus, by bringing together aforementioned techniques, we propose an \emph{agentic RAG framework with constitutional safeguards} tailored to ALICE: a curated RAG toolchain over ALICE sources; a ReAct agent exposing least-privilege tools for DCS/O$^2$ queries and routine diagnostics; and a safety/authorization layer combining constitutional policies, programmable guardrails and RBAC -- bringing state-of-the-art LLM reasoning to operational practice with traceable grounding, least-privilege access and auditability. The contribution of this work is primarily on the systems and integration side: adapting these methods to an air-gapped, safety-critical SCADA/DCS environment with explicit governance, deployment and evaluation constraints rather than proposing new base-model algorithms.

%
\section{System Design and Configuration}
\begin{figure}[ht]
  \centering
  \includegraphics[width=0.40\linewidth]{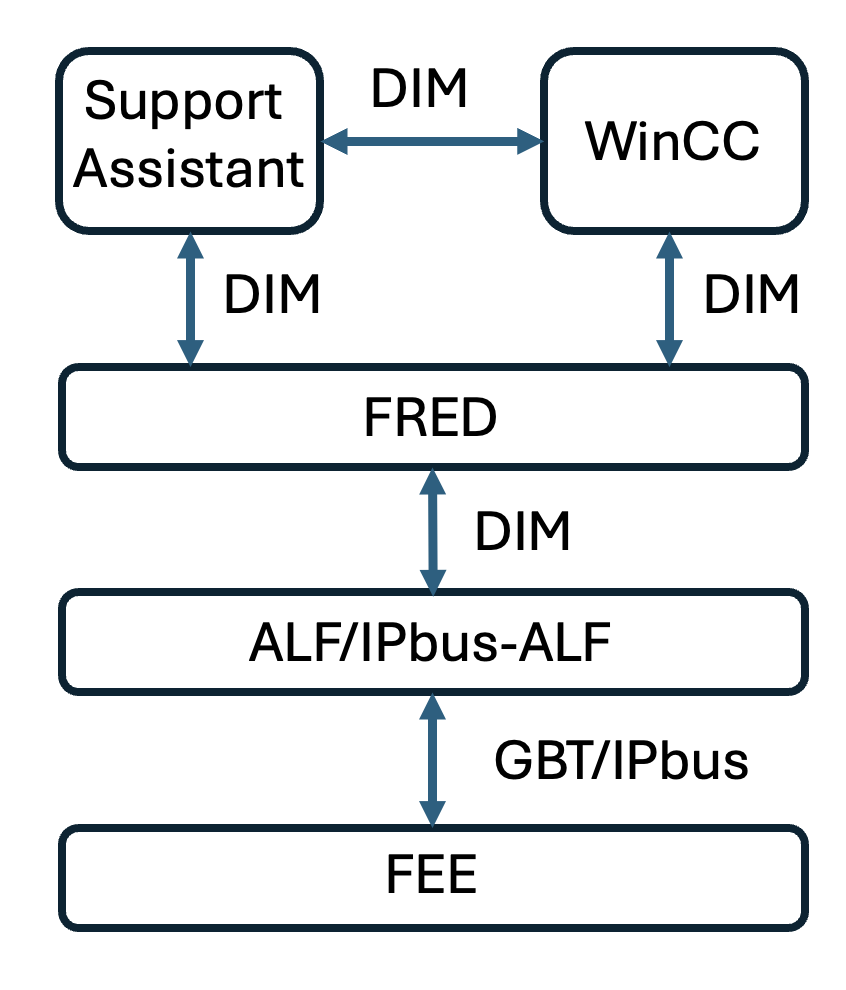}
  \caption{Control and monitoring architecture. \textit{AI-Based Support Assistant} and \textit{WinCC} exchange data bidirectionally via DIM and connect to \textit{FRED}. \textit{FRED} bridges to the \textit{ALF/IPbus--ALF} layer over DIM, which interfaces with the front-end electronics through GBT/IPbus.}
  \label{fig:control-architecture}
\end{figure}

The DCS architecture for the ALICE experiment is fundamentally based on the ALFRED (ALICE Low-Level Front-End Device) \cite{Roslon:2025zee} platform. As shown in the Figure \ref{fig:control-architecture}, this architecture consists of two primary layers. The lower layer is IPbus-ALF, a specialized implementation of the ALF framework responsible for direct communication with the detector's FEE, but instead of GBT (GigaBit Transceiver) protocol \cite{GBT}, the IPbus protocol \cite{Roslon:2025zee} may be used. The higher layer is FRED, which acts as a supervisory system. FRED processes requests, sends read/write operations to ALF which are in the next step handle on FPGA registers, and publishes the resulting data to the SCADA system.

The architecture is completed by the SIMATIC WinCC OA SCADA system, which provides the user interface for detector experts and manages communication with external components (e.g. VME and HV crates) using standard industrial protocols like OPC UA, S7 and TCP/IP. All communication between these layers is managed by the DIM (Distributed Information Management) system \cite{DIM}, a client-server middleware that relies on a DNS server for service discovery and messaging.

To successfully integrate an AI-based assistant, the system must have comprehensive access not only to low-level electronics data but also to the high-level state of all related components. This includes the Wiener VME crate powering the FEE, the CAEN high-voltage supply, and central experiment systems like the FSM, LHC and ALICE status monitors. Therefore, the proposed architecture enables the AI assistant to ingest data from both the FRED services and the WinCC OA environment, ensuring a complete operational picture.

The proposed AI assistant is designed as an independent service that interfaces with the existing ALFRED ecosystem. It is shown in the upper part at the Figure~\ref{fig:control-architecture}. It will act as both a client and a server within the DIM network, allowing it to consume real-time data and publish actionable insights.

The primary role of the AI assistant is to provide intelligent and active support to operators. Its integration into WinCC OA will focus on transforming raw data into clear, actionable guidance. The assistant autonomously detects anomalies, classifies their severity, and provides guided recovery procedures with automated expert notifications (SMS, e-mail, or JIRA tickets).
In addition to shifter-level alarms, a dedicated expert interface will be prepared. This panel will allow experts to:
\begin{itemize}
    \item \textbf{Query the AI System}: Interact with the AI to get detailed diagnostics and verify the state of the detector.

    \item \textbf{Review AI Reasoning}: Understand the data and logic that led to a specific recommendation.

    \item \textbf{Access Instructions}: Quickly find official procedures and documentation related to a given problem.
\end{itemize}
To ensure robustness and maintainability, the AI assistant will be fully integrated with CERN's standard IT infrastructure.
\begin{itemize}
    \item \textbf{Oracle Database Integration}: The AI will connect to the Oracle database within the AliDCS network to access historical data saved by FRED and FEE settings. This will allow it to perform trend analysis and make context-aware recommendations.
\end{itemize}

The pilot AI-Based support assistant shall be deployed within a dedicated FIT DCS environment, operating in a logically isolated network with no Internet connectivity. This setup ensures full compliance with CERN’s computer security policies and eliminates any external communication risks; however, it also requires that the entire AI model and its dependencies be fully deployed and maintained locally within this environment.

Access control is managed through CERN Single Sign-On (SSO) and e-group based authorization, inheriting the existing role hierarchy from the WinCC OA RBAC model. The assistant performs only read-only operations on selected telemetry, alarm logs, and documentation sources, with all actions logged for auditability.

All services, including the RAG index and inference engine, shall be hosted on ALICE DCS network under the Computer Security Baseline (CSB) framework. This setup ensures a secure, traceable, and standards-compliant environment for evaluating the assistant within the FIT control system.

\subsection{Knowledge-Base Governance and Corpus Curation}

To prevent stale, inconsistent, or overly sensitive information from influencing recommendations, the retrieval corpus is governed by an explicit curation policy. Only sources that are already part of the official ALICE operational toolchain (e.g.\ approved procedures, JCOP/WinCC~OA documentation, selected Bookkeeping and eLog entries) are considered for inclusion. New documents enter the corpus through a controlled pipeline in which FIT experts assign a document type, subsystem tags, and a validity interval; when a procedure is superseded, the corresponding entry is either retired or explicitly marked as obsolete. 

The RAG index is built from versioned snapshots of this corpus. Each retrieved chunk carries a document identifier, version, and timestamp, which are surfaced to the operator together with the assistant's answer. Redaction rules allow removal or masking of information that should not be exposed at the AI layer (for example, credentials, hostnames, or low-level configuration details that are not relevant for shifters). Periodic re-indexing ensures that corrections in the underlying documents are propagated in a controlled way, and an audit trail of corpus changes is maintained to support post-mortem analyses of recommendations.

\subsection{From Read-Only Advice to Safe Action}

In the initial pilot, the assistant operates in pure read-only mode: it can query telemetry, alarm histories and documentation, but it cannot execute any control action. As the system matures, a restricted set of \emph{Quick Commands} will be exposed as WinCC~OA buttons backed by the AI service. Each such command corresponds to a pre-approved, parameter-bounded procedure (e.g.\ resetting a specific subset of front-end boards under known safe conditions) and is implemented as a separate, testable endpoint.

Safety gating uses a multi-layered approach. First, role-based access control ensures that only users with appropriate privileges can see or trigger certain commands. Second, the assistant never initiates write actions autonomously: instead, it proposes a command together with a natural-language justification and a preview of the concrete low-level operations, and the operator must explicitly confirm execution in the WinCC~OA panel. Third, the command service validates preconditions against the current FSM state and equipment status; if any safety constraint is violated (for example, an unexpected beam condition or a conflicting alarm), the action is blocked and the operator is informed. In this way, the ReAct-style tool use is constrained by both human confirmation and deterministic checks in the control stack.




\section{Implementation}

The plan begins with a foundational prototype to establish core connectivity and progressively builds towards a fully-featured, integrated support system. The initial and most critical phase is the development of a working prototype. The primary goal of this stage is to establish a complete, end-to-end communication path between all system layers to validate the architecture and de-risk subsequent development.

The prototype will focus on:
\begin{itemize}
    \item \textbf{Establishing the Communication}: Creating the bidirectional data flow:
    \begin{enumerate}
        \item \textit{Direct FRED Data Publishing}: The AI assistant will subscribe to data streams from FRED and the high-level context from WinCC OA.
        \item \textit{SCADA and Central Systems Context}: The AI assistant will publish its initial analyses (e.g., a system "health" status).
        \item \textit{SCADA Reception:} The WinCC OA system will subscribe to the AI's data flow and display the received information in a dedicated panels integrated with FSM and ALICE central systems.
    \end{enumerate}
    \item \textbf{Integration with External Tools}: The prototype will also include the necessary components for connecting to essential external systems, such as the Oracle database for accessing historical data and Grafana for monitoring the AI assistant's own performance metrics.
\end{itemize}

At the application level, the assistant is implemented as a ReAct-style agent that iteratively reasons over the current context, decides whether to call a tool, incorporates the observation, and updates its plan. Tools include time-bounded queries over DCS telemetry, retrieval calls into the curated document index, and accessors for recent alarm and FSM histories. Anomaly detection leverages both simple statistical baselines (for example, monitoring deviations of key sensor readings from rolling medians) and rule-based patterns derived from existing alarm logic. Detected anomalies are grouped into incidents and labeled with a coarse severity class (\emph{info}, \emph{warning}, \emph{critical}) using a supervised classifier trained on historical incidents and expert labels; the LLM then uses this severity information to prioritize and phrase its recommendations but does not change the underlying classification.

With the communication infrastructure in place, the next phase focuses on developing the core features that provide direct value to the operators.

\begin{itemize}
    \item \textbf{Proactive Alerts and Checklists}: The AI will be programmed to continuously monitor the incoming data streams for predefined anomaly signatures. Upon detection, it will generate proactive alerts that include contextual information.
    
    \item \textbf{Quick Commands}: To streamline operator workflows, the assistant will offer context-aware, single-click actions presented within the WinCC OA interface that execute common procedures recommended by the AI (e.g., a "Reset System" button).
\end{itemize}

To deliver a seamless user experience, the integration with WinCC OA will leverage its advanced capabilities and a dedicated software layer to manage communication. Advanced UI components are embedded using WinCC OA’s EWO framework \cite{EWO}, with a dedicated asynchronous manager ensuring responsive communication with the AI.

The implemented functionalities will be validated against a set of predefined operational scenarios, including:
\begin{itemize}
    \item \textbf{Error Diagnostics}: Guiding an operator through identifying the root cause of common faults.
    \item \textbf{Status Monitoring}: Providing a high-level, "at-a-glance" summary of the detector's health.
    \item \textbf{Action and Escalation Recommendations}: Suggesting appropriate corrective actions and advising when to escalate a problem to an expert.
\end{itemize}
\section{Testing and Evaluation}
\begin{figure}[ht]
  \centering
  \includegraphics[width=\linewidth]{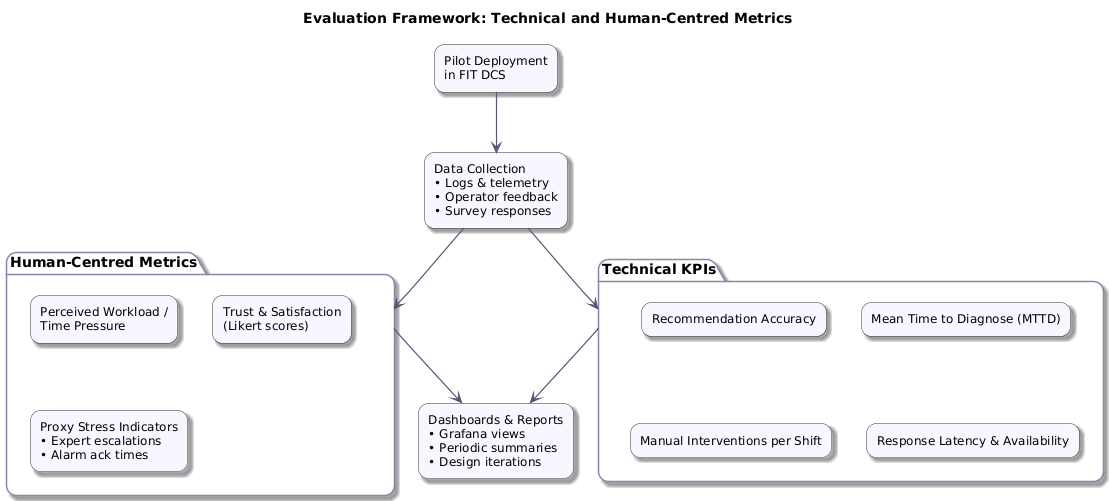}
  \caption{Overview of the evaluation framework. The methodology integrates quantitative technical KPIs (e.g., accuracy, MTTD) with qualitative human-centred metrics (e.g., perceived workload, trust) to provide a holistic assessment of the AI assistant's impact on detector operations.}
  \label{fig:evaluation-framework}
\end{figure}

Before deployment on the FIT system, the AI Support Assistant will be tested in a laboratory environment to verify functionality, response quality, and system stability under controlled conditions. Following successful validation, it will be integrated into the FIT DCS pilot, where tests will focus on answer accuracy, response latency, and long-term operational stability during routine control-room operation. General flow of the evaluation setup is visualized in Figure~\ref{fig:evaluation-framework}.

In the initial testing phase, the assistant will not perform any direct control actions within the DCS; its role will be limited to providing contextual assistance and actionable suggestions to the operator. The evaluation will be carried out together with FIT DCS experts to ensure reliability, factual correctness, and compliance with established operational procedures.

The AI assistant's resource consumption (CPU, memory), application logs, and overall status will be continuously monitored and visualized in a dedicated Grafana dashboard, providing developers with full transparency into its performance.

To objectively measure the assistant's effectiveness, a set of Key Performance Indicators (KPIs) will be continuously tracked. These metrics will guide the development of the AI's alert and recommendation models and will be also visualized in Grafana. The evaluation framework is structured around several core KPIs. Additionaly, as illustrated in Fig.~\ref{fig:evaluation-framework}, the assessment strategy distinguishes between technical performance metrics and human-centred factors to ensure the system supports operators effectively without increasing cognitive load.

\begin{itemize}
    \item \textbf{Recommendation Accuracy Rate (\%)}: This metric assesses the correctness of the AI's suggestions.
    \begin{itemize}
        \item \textit{Data Collection}: A feedback mechanism will be integrated into the WinCC OA alert panel, allowing operators to classify each AI suggestion as "Correct," "Partially Correct," or "Incorrect."
        \item \textit{Visualization}: A Grafana pie chart will display the feedback distribution, while a time-series graph will track the accuracy rate over the run period.
    \end{itemize}
    \item \textbf{Mean Time to Diagnose (MTTD)}: This metric measures the assistant's impact on accelerating troubleshooting, with the time measured in minutes.
    \begin{itemize}
        \item \textit{Data Collection}: The system will automatically log timestamps for the initial anomaly trigger and subsequent operator actions.
        \item \textit{Visualization}: A histogram will compare the MTTD for incidents handled with and without the AI's assistance.
    \end{itemize}
    
    \item \textbf{Number of Manual Interventions per Shift}: This KPI quantifies the reduction in operator workload.
    \begin{itemize}
        \item \textit{Data Collection}: All operator-initiated commands from the WinCC OA interface will be logged and cross-referenced with the AI's recommendations.
        \item \textit{Visualization}: A time-series graph will plot the number of manual interventions, aiming to show a decreasing trend over time.
    \end{itemize}

    \item \textbf{Response Latency and Availability}: This KPI captures whether the assistant can answer in a time frame compatible with control-room workflows.
    \begin{itemize}
        \item \textit{Data Collection}: For each request, the end-to-end response time is logged together with the system load. We target a steady-state 95th-percentile (p95) latency below a few seconds for typical single-shifter usage and monitor uptime of the assistant service.
        \item \textit{Visualization}: Grafana will display latency distributions (p50, p95) and availability over time, highlighting periods in which performance targets are not met.
    \end{itemize}

    \item \textbf{Human-Centred Metrics}: These indicators assess how the assistant affects the human operator.
    \begin{itemize}
        \item \textit{Data Collection}: During the pilot, shifters will be invited to fill short questionnaires before and after the introduction of the assistant and at regular intervals. These will capture perceived workload, time pressure, trust in the assistant's suggestions, and overall satisfaction on a Likert scale. As an additional proxy for stress, we will monitor objective indicators such as the number of escalations to experts and the distribution of alarm acknowledgement times.
        \item \textit{Visualization}: Aggregated scores will be plotted over time and correlated with technical KPIs, allowing us to assess whether improvements in MTTD or workload reduction also translate into better operator experience.
    \end{itemize}

\end{itemize}

The pilot phase will provide feedback necessary to refine the assistant’s configuration and interaction model within the FIT DCS. After the completion of LHC Run 4, the LHC Long Shutdown 4 (LS4) period — lasting 2.5 years \cite{Roslon:2025lxb} — will offer the opportunity to adapt and generalize the system for use beyond FIT.

This phase will focus on expanding the retrieval corpus, harmonizing integration interfaces, and tuning the assistant to handle subsystem-specific diagnostics. The outcome will form the basis for a scalable, multi-detector AI support platform to be deployed for LHC Run 5.
\section{Summary and Further System Development}
The proposed AI-based assistant for the ALICE-FIT detector is intended to enhance operational efficiency and safety within the Detector Control System by providing context-aware diagnostics and proactive decision support. The system will integrate Large Language Models with a Retrieval-Augmented Generation framework to deliver verified, traceable, and explainable recommendations directly within the WinCC OA environment. It will operate in read-only mode under CERN’s Computer Security Baseline and Single Sign-On framework, ensuring compliance and auditability.
The prototype will initially be deployed in a dedicated FIT DCS environment to validate its architecture, performance, and usability in real operational scenarios. Key performance indicators—recommendation accuracy, mean time-to-diagnose, and operator workload reduction—will guide further optimization. Future development will focus on improving reasoning reliability, expanding the retrieval corpus to other ALICE subsystems, and preparing a scalable, multi-detector platform for deployment during LHC Run 5.

\vspace{2mm}

\noindent 

\vspace{1mm}

\noindent \textbf{Disclosure of Interests.} The authors have no competing interest to declare that are relevant to the content of this article.
\newline
\noindent \textbf{Funding.} This work was supported by the Polish Ministry for Education and Science under agreements
no. 5452/CERN/2023/0.

%
%
\end{document}